\documentclass{aastex}
\usepackage{spr-astr-addons}
\usepackage{url}

\RequirePackage{color}

\begin{document}

\title{Spectral atlases of the Sun from 3980 to 7100 \textup{\AA}\\ at the center and at the limb}
\slugcomment{Not to appear in Nonlearned J., 45.}
\shorttitle{Spectral atlases of the Sun} \shortauthors{Fathivavsari
et al.}

\author{H. Fathivavsari\altaffilmark{1,2}} \and \author{A. Ajabshirizadeh\altaffilmark{1,3,4}}
\and
\author{S.~Koutchmy\altaffilmark{2}}

\altaffiltext{1}{Department~of~Theoretical~Physics~and~Astrophysics,~University~of~Tabriz,~Tabriz~51664,~Iran.\\
e-mail: \textcolor{blue}{h.fathie@gmail.com}}

 \altaffiltext{2}{Universit\'e
Pierre et Marie Curie, CNRS - UMR 7095, Institut d'Astrophysique de
Paris CNRS and UPMC, 98bis Boulevard Arago, 75014 Paris, France.}

\altaffiltext{3}{Research~Institute~for~Applied~Physics~and~Astronomy,~University~of~Tabriz,~Tabriz~51664,~Iran.}

\altaffiltext{4}{Research Institute for Astronomy and Astrophysics
of Maragha (RIAAM) \& Center for Excellence in Astronomy \&
Astrophysics of Iran (CEAA), Maragha, Iran, P.O.Box: 55134-441.}

\begin{abstract}

In this work, we present digital and graphical atlases of spectra of
both the solar
   disk-center and of the limb near the Solar poles using data
taken at the UTS$-$IAP \& RIAAM (the University of Tabriz
Siderostat, telescope and spectrograph jointly developed with the
Institut d'Astrophysique de Paris and Research Institute for
Astronomy and Astrophysics of Maragha). High resolution and high
signal-to-noise ratio (SNR)
   CCD-slit spectra of the sun for 2 different parts of the disk, namely for
   $\mu$~=~1.0 (solar center) \& for $\mu$~=~ 0.3 (solar limb) are provided and discussed.
While there are several spectral atlases of the solar disk-center,
this is the first spectral atlas ever produced for the solar limb at
this spectral range. The resolution of the spectra is about
\emph{R}~$\sim$~70 000  ($\Delta\lambda$~$\sim$~0.09~\textup{\AA})
with the signal-to-noise ratio (SNR) of 400$-$600. The full atlas
covers the 3980 to 7100 \textup{\AA} spectral regions and contains
44 pages with three partial spectra of the solar spectrum put on
each page to make it compact. The difference spectrum of the
normalized solar disk-center and the solar limb is also included in
the graphic presentation of the atlas to show the difference of line
profiles, including far wings. The identification of the most
significant solar lines is included in the graphic presentation of
the atlas. Telluric lines are producing a definite signature on the
difference spectra which is easy to notice. At the end of this paper
we present only two sample pages of the whole atlas while the
graphic presentation of the whole atlas along with its ASCII file
can be accessed via the ftp server of the CDS in Strasbourg via
anonymous ftp to cdsarc.u-strasbg.fr (130.79.128.5) or via this link
\footnote{
\url{http://cdsarc.u-strasbg.fr/viz-bin/qcat?J/other/ApSS}}.

\end{abstract}



\section{Introduction}

The existing solar atlases can be divided into two main categories:
1) solar full disk atlases (a.k.a solar flux or even irradiance
atlases); 2) solar disk-center spectral atlases. A pioneering work
of full disk solar atlases was the solar irradiance spectrum of Labs
\& Neckel (1968). This atlas covers a wide range of wavelengths but
has a rather low spectral resolution such that even the major
Fraunhofer lines were not resolved. It was not until 1976 that J.
Beckers and his collaborators (Beckers et al. 1976) published the
first high resolution solar flux atlas which covers the wavelength
range of 3800 $-$ 7000~\textup{\AA}. Later on, Kurucz et al. (1984)
used spectra taken at the McMath-Pierce Solar Telescope located on
Kitt Peak Observatory, to publish their high resolution and high SNR
solar flux atlas, covering the 2960 $-$ 13000~\textup{\AA} spectral
region.

\begin{figure*}
\begin{center}
\resizebox{11.5cm}{!}{\includegraphics{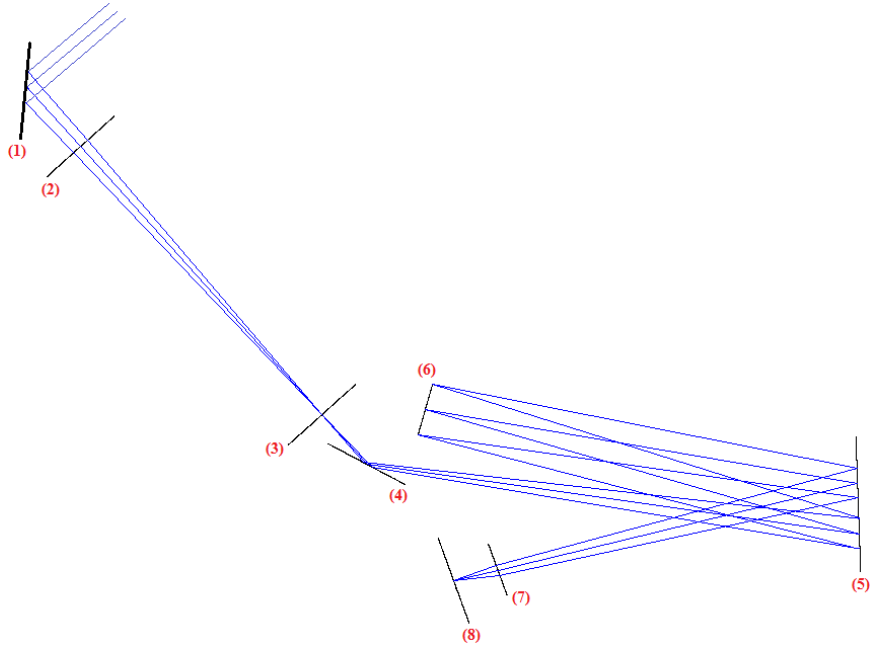}} \hfill
\caption{The optical layout of the instrument: 1) Siderostat primary
35~cm diameter plane mirror; 2) Zeiss astrograph F/15 objective with
focal length of 3~m; 3) Adjustable slit; 4) Fold mirror 5) F/12
collimator with 30~cm diameter; 6) 16 cm Grating with 1800
grooves/mm and $\lambda_{blaze}$~$=$~5500~\textup{\AA}; 7) Lens with
20~cm focal length; 8) The CCD.}
\end{center}
\end{figure*}

The first widely used high resolution disk-center slit solar
spectrum known as the Utrecht Spectrum was produced from
photographic plates following a large effort by Minnaert et al.
(1940) and many others. The original spectrograms were obtained at
the 150- foot tower telescope of Mount Wilson in 1936-37. This atlas
was later analyzed by Moore et al. (1966). Furthermore, the Liege
and Jungfraujoch central intensity atlas, \emph{Spectrophotometric
Atlas of the Solar Spectrum from $\lambda$3000 to $\lambda$10000}
was published by Delbouille et al (1973). To obtain their
photo-electric spectra recorded with several photo-multipliers to
cover the whole spectral range, they repeatedly used a rapidly
scanning large spectrometer fed by a coelostat at the high altitude
Jungfraujoch station and selected the best digital recordings. This
atlas is available by parts from the French BASS2000\footnote{
\url{http://bass2000.obspm.fr/solar_spect.php}} ground-based solar
data base. More recently, Wallace et al. (1998) published an atlas
of the solar spectrum at disk center, based on archival Fourier
transform spectrometer (FTS) data taken at the McMath-Pierce Solar
Telescope. Several attempts were also made by Amateurs to produce a
compact atlas more limited in quality but more practical \footnote{
\url{http://www.astrosurf.com/spectrohelio/spectre_solaire-en.php}}.

\begin{figure}
\begin{center}
\resizebox{5.5cm}{!}{\includegraphics{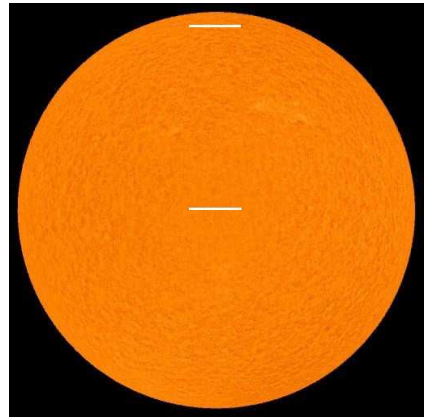}} \hfill
\caption{The positions of the slit on the solar disk. The two while
lines show the positions of the slit at the solar disk-center
($\mu$~=~1.0) and on the solar limb ($\mu$~=~0.3).}
\end{center}
\end{figure}

In the current work, we present an atlas of the solar disk-center
($\mu$~=~1.0) of the quiet Sun along with that of the solar limb
($\mu$~=~0.3) and make a comparison. Spectra are based on data taken
at the University of Tabriz Siderostat developed in cooperation with
the Institut d'Astrophysique de Paris (UTS$-$IAP) in 2000 - 2005.
Noting that very large telescopes today are capable of producing
rather good resolution spectra, even for far objects, we hope this
atlas along with its companion ASCII file would contribute to
researches in solar, interplanetary, cometary, stellar, galactic and
extra-galactic physics. We note that this atlas is the first of a
series of atlases. The future atlases will cover the near-infrared
part of the spectrum and even more importantly and more difficult to
produce spectra of the core of a solar sunspot (work in progress).

\section{Set-up and Observations}

To make the current atlas we use the UTS-IAP solar telescope and
spectrograph (see the optical layout of the instrument in Fig.~1).
The grating has 1800 grooves per millimeter with the blaze
wavelength at 5500~\textup{\AA}. Our spectrograph is in the
Ebert-Fastie configuration and the 30 cm diameter F/12 collimator is
used for both collimating and focusing the incident light. As
illustrated in Fig.~1, the light of the sun enters the spectrograph
via the objective lens, after first striking the siderostat primary
35 cm diameter plane mirror. The focal length of the Zeiss
astrograph F/15 objective is 3~m, and the slit width of the
spectrograph is fixed at 20~$\mu$m which gives us a resolution of
\emph{R}~$\sim$~70~000. The sunlight passes through the slit and
strikes the collimator located at 3.7~m away. The collimated beam of
light then hits the grating which is also 3.7~m away from the
collimator. The grating disperses the light into a 1st order
spectrum and then reflects it back on the collimator. The collimator
is now focussing the dispersed light on the CCD chip. We emphasize
that, in order to increase the wavelength coverage of the CCD chip,
we equipped the CCD with a lens of 20~cm focal length to slightly
demagnify (see below). Table~1 reports the date of observation,
solar altitude and humidity at the time of observations.

\begin{figure*}
\begin{center}
\resizebox{17.0cm}{!}{\includegraphics{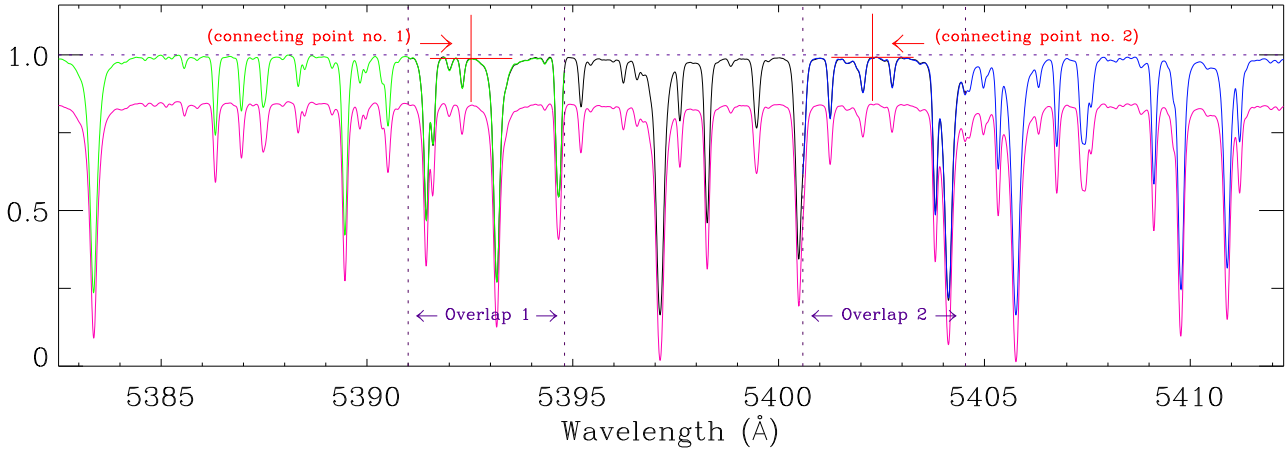}} \hfill
\caption{An illustration of the procedure used to juxtapose all
individual observed spectral windows. The green, black and blue
curves are the three spectral windows that are to be combined. The
two red crosshairs mark the positions of the two connecting points
and the pink curve is the final combined spectrum. For the sake of
comparison, the final spectrum (i.e. the pink curve) is shifted
along the y-axis.}
\end{center}
\end{figure*}

\begin{figure*}
\begin{center}
\resizebox{17.0cm}{!}{\includegraphics{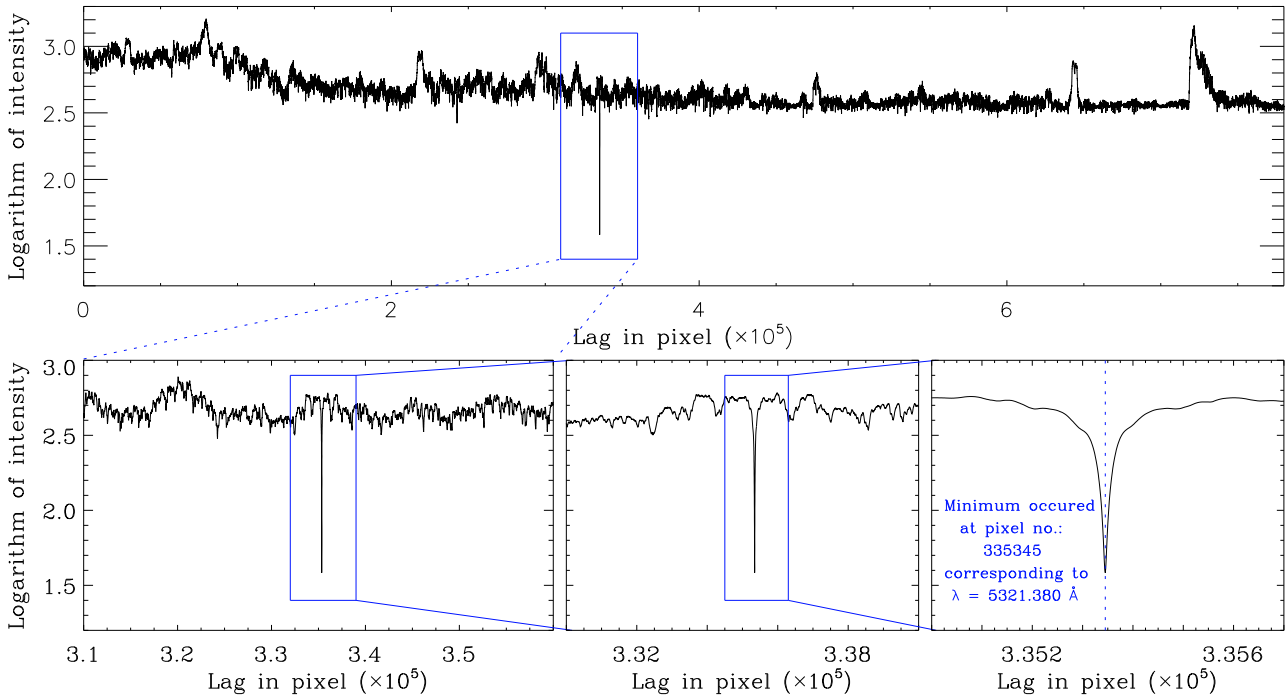}} \hfill
\caption{A cross-correlation of an observed spectral window and the
KPNO spectrum. The position of the minimum corresponds to the lag at
which the two spectra overlap the most. This lag is then used to do
wavelength calibration on the observed spectrum.}
\end{center}
\end{figure*}

We used an Atik ATK-2HS CCD camera to record the observed
2$-$dimensional (2D) spectral images. The CCD chip is of
640$\times$480 pixels, that is 640 pixels along the dispersion axis
and 480 pixels along the slit. The width and the length of the slit
covers $\sim$~1.4 and $\sim$~75 arcsecond on the solar surface,
respectively. Sunspots and pores were avoided and no faculae were
seen at the polar limb during the used recordings. Fig~2 shows the
two different positions of the slit on the solar disk. In this
figure, the two white lines indicate the positions of the slit. In a
single exposure, the 640 pixels along the dispersion axis could only
cover 11 to 15 angstroms of the solar spectrum. Therefore, we
required more than 330 separate 2D spectral images to cover the
whole spectral range of interest. These spectral segments are
extracted and then assembled to form the final atlas. We note that
each of these 2D images has about 3 to 5 angstrom overlap with its
successive images. These overlaps are then used to juxtapose all the
individual exposures in order to make one single spectrum that
ranges from 3980 to 7100~\textup{\AA}. To this end, each spectral
window is plotted with its two neighboring windows over-plotted on
it with different colors (see Fig.~3). We then interactively mark
two appropriate connecting points (away from the troughs of the
absorption lines), one for each neighboring window. The three
successive spectra are then combined by averaging the intensity of
each of the two overlapping windows at their corresponding
connecting points. This is done for all the observed spectral
windows. In Fig.~3, the two red crosshairs indicate the positions of
the two connecting points. The final combined spectrum is also shown
as a pink curve. Note that for the sake of comparison, the y-axis of
the final combined spectrum (i.e. the pink curve) is shifted along
the y-axis.

As for the Atlas (called KPNO hereafter) we compare our work with,
the spectrum (Wallace et al. 1998) was recorded using a Fourier
Transform Spectrometer (FTS; Davis et al. 2001). The solar image
from the large Kitt-Peak siderostat tower telescope of 1.6~m
aperture is focused on a rotating vertical table at the entrance
aperture of the FTS. These apertures are circular and their diameter
ranges from 0.5~mm (1.2~arcsec) to 10~mm (25~arcsec). The final
spectrum is then obtained by computing the Fourier transform of the
recorded interferograms. It usually takes 3 to 14 minutes to
complete a single, full resolution scan. Several of such scans are
generally taken in order to improve the SNR of the interferograms.

 \begin{table}
\caption{Observing condition at the time of observation. Note that
the observations were done in 2013 in rather different conditions
for Earth atmospheric transmission.}
\setlength{\tabcolsep}{2.5pt}
\begin{tabular}{c| c c c}
\hline\hline

     & Date of Obs.  & Solar altitude~  & Humidity                   \\
\hline

Solar Center   &    June 7 \& 8             & 70~$^{o}$ & 30~\%      \\
Solar Limb     &    Nov.  18 \& 19           & 30~$^{o}$ & 55~\%                    \\

 \hline
\end{tabular}
\renewcommand{\footnoterule}{}
\end{table}

\section{The Wavelength Calibration}

Each spectral window was wavelength-calibrated separately. We used a
new method for doing the wavelength calibration. What is usually
done in spectroscopy is to take spectra of calibration lamps (ex.
Th-Ar; He-Ne lamps) with the same setup that the object spectra were
obtained with. However, if obtaining on-board calibration data is
not possible (ex. in the case of spectra recorded by SUMER
spectrograph mounted on SoHO space telescope) wavelength calibration
is done with the help of some stable photospheric spectral lines
(Vial et al. 2007). But in the current work, we use
cross-correlation analysis in order to do wavelength calibration on
our observed spectra. Cross-correlation analysis is the most popular
procedure for deriving relative velocities between an observed
stellar spectrum and a template spectrum (Tonry \& Davis 1979). We
note that, in this work, the solar spectrum observed by Wallace et
al. (1998) (i.e. KPNO spectrum) is used as the template spectrum for
our cross-correlation analysis.

A cross-correlation measures the similarity between two signals. For
this purpose, the whole KPNO spectrum is searched for the pattern
similar to that of the observed spectral window. To this end, the
observed spectrum is shifted iteratively with respect to the KPNO
spectrum by a series of regularly spaced increments. We then compute
the sum of the normalized flux differences of the shifted and
unshifted spectrum at each pixel. The output of the
cross-correlation between one sample observed spectrum and the KPNO
spectrum is shown in Fig.~4. In this figure, the position of the
cross-correlation minimum (i.e. pixel no. 335345) corresponds to the
lag between the two spectra.

As noted in sect.~2, our CCD has 640 pixels along the dispersion
axis while the KPNO spectrum, for the same wavelength range,
contains $\sim$~1000 (in the red part of the spectrum) up to
$\sim$~5000 (in the blue part of the spectrum) data-points.
Therefore, before cross-correlating the two spectra, each observed
spectral window is re-sampled to match the KPNO spectrum wavelength
grid.

\section{Instrumental profile}

Telluric lines, especially those of the molecular oxygen bands are
much narrower than solar absorption lines, and their profiles can
provide an indication of the instrumental profile of the
spectrograph (Griffin 1969). Fig.~5 presents the absorption profiles
of the telluric $\alpha$~-band (upper panel) and B-band (lower
panel) of oxygen taken from the current atlas. To distinguish
between the telluric and solar absorption in this region, we put a
blue vertical tick below each solar absorption line. In the upper
panel of Fig.~5, the two lines marked by the green stars are the two
absorption used in determining the instrumental profile of the
spectrograph (see below).

\begin{figure*}
\begin{center}
\resizebox{17.0cm}{!}{\includegraphics{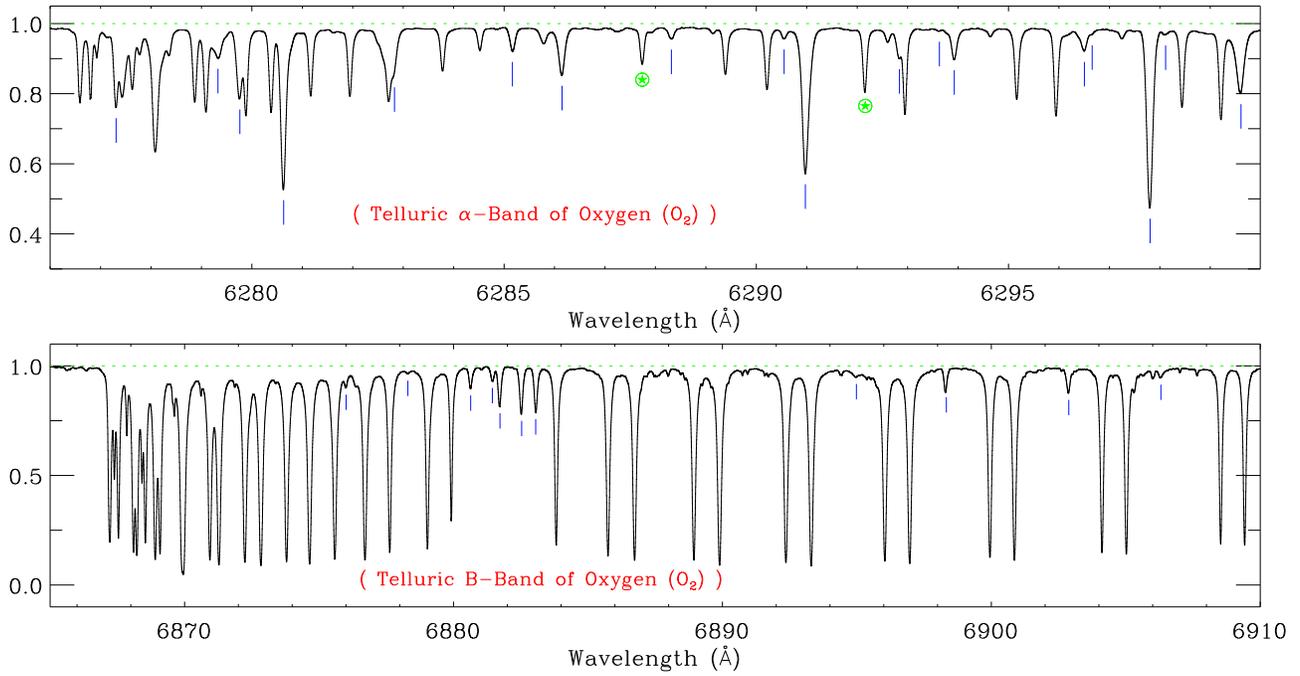}} \hfill
\caption{The absorption profiles of the telluric $\alpha$~-band
(upper panel) and B-band (lower panel) of oxygen taken from the
current atlas. The absorption lines marked by the blue vertical
ticks are solar. In the upper panel, the two lines marked by the
green stars are the two absorption lines used in determining the
instrumental profile of the spectrograph.}
\end{center}
\end{figure*}

\begin{figure*}
\begin{center}
\resizebox{12.0cm}{!}{\includegraphics{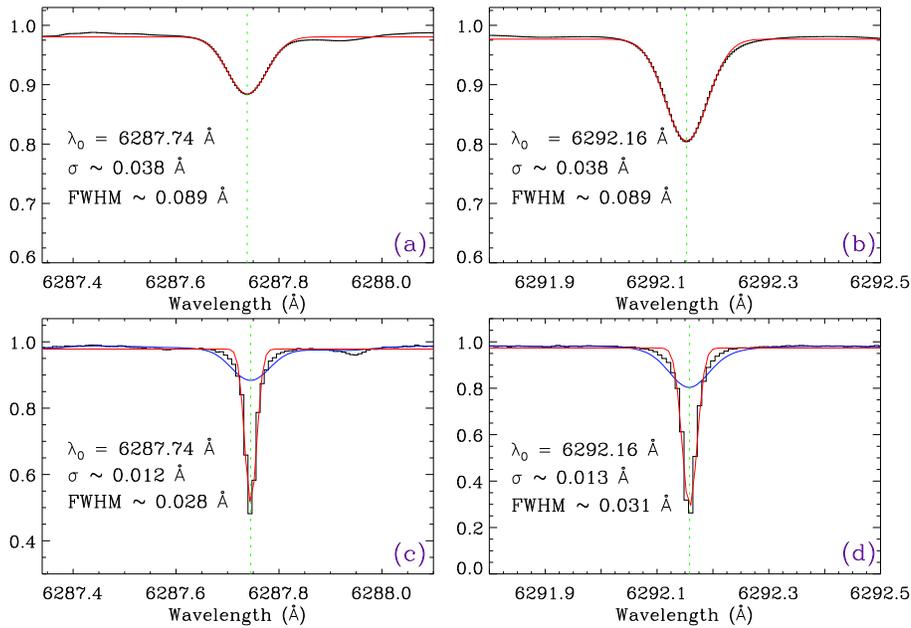}} \hfill
\caption{The two absorption lines of the telluric $\alpha$~band of
oxygen with $\lambda$~$\sim$~6287.74~\textup{\AA} and
$\lambda$~$\sim$~6291.16~\textup{\AA} are used to compare the
instrumental profiles of the current atlas (panels (a) and (b)) with
that of the Wallace et al. (1998) (panels (c) and (d)). A Gaussian
fit (red curve) is conducted on each of the individual absorption
line (black curves) to determine the Gaussian width ($\sigma$) as
well as the FWHM of the line. Note that in panels (c) and (d), for
the ease of comparison, the corresponding telluric lines of the
current atlas are also shown as blue curves.}
\end{center}
\end{figure*}

We compare, in Fig.~6, the instrumental profile of the current atlas
with that of the KPNO spectrum. For this purpose, we conducted
Gaussian fits (red curves) to two absorption lines (black curves) of
the telluric $\alpha$ band of Oxygen, namely, the ones at
$\lambda$~$\sim$~6287.74~\textup{\AA} and
$\lambda$~$\sim$~6291.16~\textup{\AA}. The upper two panels in
Fig.~6 (i.e. panels (a), and (b)) are for the current atlas while
the lower two panels belong to the KPNO spectrum. Moreover, in each
panel, we report the central wavelength of the profile
($\lambda_{0}$), the Gaussian width ($\sigma$), and the FWHM of the
absorption lines. Note that in panels (c) and (d), for the ease of
comparison, the corresponding telluric lines of the current atlas
are also shown as blue curves.

The FWHM of the two telluric absorption lines of the current atlas
(panels (a) and (b) in Fig.~6) are $\sim$~0.09~\textup{\AA},
resulting in a resolving power of
\emph{R}~=~$\lambda$/$\Delta\lambda$~$\sim$~70~000. At this
\emph{R}, the spectral resolution bin of $\Delta\lambda$~$\sim$~0.09
is sampled by four detector pixels (i.e. each pixel of 7.4~$\mu$m
length covers $\sim$~0.02~\textup{\AA}.). On the other hand, in the
KPNO spectrum (panels (c) and (d)), the FWHM of
$~$$\sim$~0.03~\textup{\AA} yields a resolving power of
$~$$\sim$~210 000.


\begin{figure*}
\begin{center}
\resizebox{17cm}{!}{\includegraphics{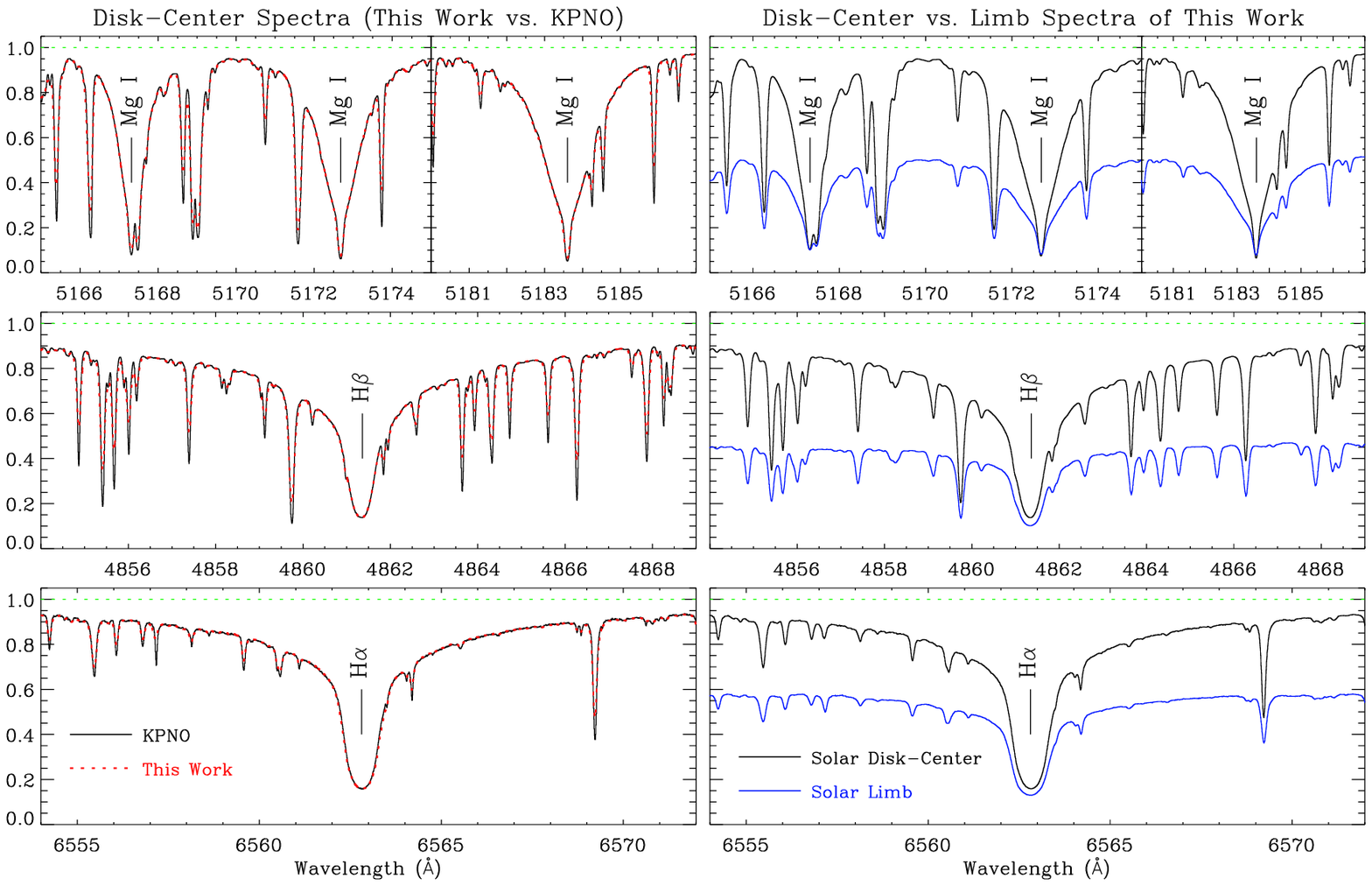}} \hfill
\caption{Comparing the line profiles of H~$\alpha$, H~$\beta$, and
Mg~I triplet of the KPNO spectrum and those of the current work. In
the left-hand side panels, the solar disk-center spectrum of the
KPNO (black curves) is compared with that of the current work (red
curves). In the right-hand side panels, on the other hand, our solar
disk-center (black curves) and solar limb (blue curves) spectra are
compared. We note that, in the case of the limb spectra, the level
of the continuum was normalized to the values given in Allen
(1981).}
\end{center}
\end{figure*}


\begin{figure*}
\begin{center}
\resizebox{10.cm}{!}{\includegraphics{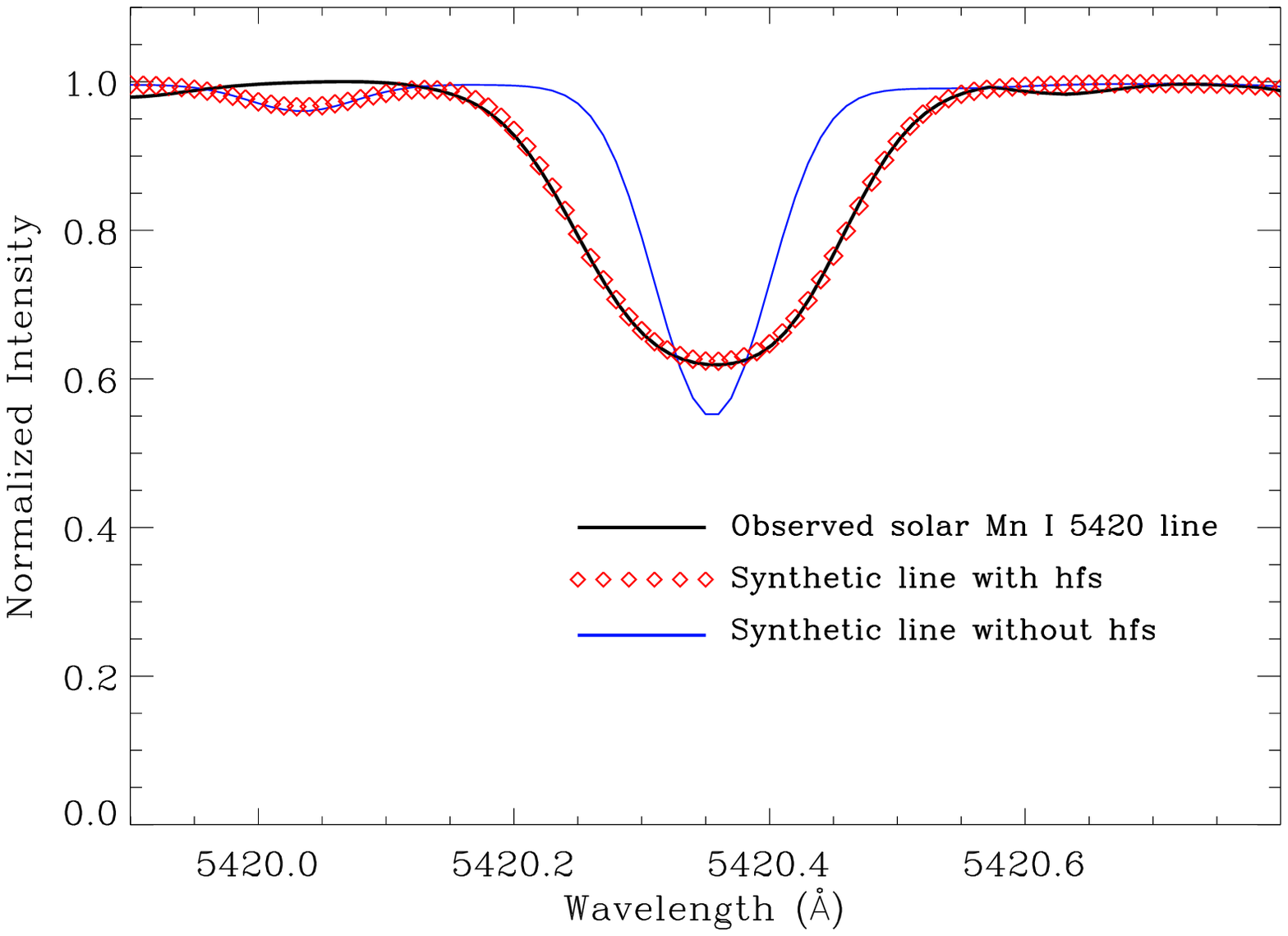}} \hfill
\caption{A demonstration of the effect of hyperfine structure on the
Mn~{\sc i} absorption profile observed with the UTS$-$IAP
spectrograph. This is the same as Figure~3.1 of Gray (2010) but
reproduced using our own observed solar spectrum.}
\end{center}
\end{figure*}

\section{Discussion}

The spectra used for the current work were taken at the solar
disk-center with $\mu$~=~1.0 and the solar limb with $\mu$~=~0.3. We
tried to avoid plage, sunspots and active regions during our
observations. The quality of the spectrum is high as the achieved
SNR is 400-600, thanks to the integration performed along the slit.
The final atlas, ranging from 3980 to 7100 \textup{\AA}, contains 44
pages of graphic presentation of the spectrum with the
identification of most of the lines included. Each page contains 3
partial spectra. The identification of lines is made using the
reference atlas of Moore et al. (1966). Two sample pages of the
whole atlas are shown at the end of this paper. Each page contains
three spectral windows, with each window made up of two panels. The
upper panel shows the spectrum of the solar disk-center (black
curve) on which the spectrum of the solar limb is overplotted as the
red curve while the lower panel illustrates the difference spectrum
of the two (i.e. I($\mu$~=~1.0)~$-$~I($\mu$~=~0.3)). The difference
spectrum of the lower panel clearly indicates that the absorption
lines taken at the solar disk-center are usually deeper than those
taken at the solar limb (in this panel the two pink dashed lines
indicate the residual intensity of 0.05 above and below zero). Note
that, in the solar limb spectrum, the level of the continuum was put
to the reference level given in the \emph{Astrophysical Quantities,
3rd Edition} by C. W. Allen, p. 169. Moreover, the ASCII file of the
atlas is also provided as a companion to this paper.

In Fig.~7, the absorption profiles of H~$\alpha$, H~$\beta$, and
Mg~I triplet of the KPNO spectrum and current atlas are presented
and compared. In this figure, in the left-hand side panels, the
solar disk-center spectrum of the KPNO (black curves) is compared
with that of the current work (red curves). In the right-hand side
panels, on the other hand, our solar disk-center (black curves) and
solar limb (blue curves) spectra are compared. As illustrated in
this figure, the wings of the Balmer lines (i.e. H~$\alpha$ and
H~$\beta$) are narrower at the solar limb than they are at the solar
disk-center.

In central ($\mu$~=~1) intensity, the continuum is formed at about
5500~K while in the case of $\mu$~=~0.3, we do not see as far in,
instead, we only see down to 5000~K or so where the continuum is
less bright (Vernazza et al. 1981). Moreover, the line wings are
produced by Stark broadening which is proportional to the electron
number density (Gray 2008). Inspecting the Kurucz Solar atmosphere
model (Kurucz 1993) at T~$\sim$~5500 and 5000~K shows that the
electron number density is lower for the latter temperature.
Therefore, the lower electron number density can result in a lower
amount of Stark broadening, and hence narrower wings, as seen for
the Balmer lines in Fig.~7. In this figure, the center-limb
variations of intensities of the Balmer lines show a limb darkening
effect whatever the position in the line, as has been established
for a long time (Dunn, 1959; White, 1962). We note that, in Fig.7,
in the case of the limb spectra, the level of the continuum was
normalized to the values given in Allen (1981).

As shown in Table~1, during the solar limb observation, the humidity
was higher than when the solar disk-center data were being recorded.
As a result, the features seen in the difference spectrum for the
telluric lines are all above zero, making them easy to identify.
Needless to say, the lower altitude of the sun during the solar limb
observation also had some contribution to this effect.

Finally, we want to point out that the quality of our compacted
digital spectra is sufficient to make a research quality work using
the recorded line profiles. The width of chromospheric lines is
readily recognized. Indeed, there are several photospheric
absorption lines in the observed solar spectrum that are unusually
broad; a simple Gaussian or Voigt profile fit cannot reproduce the
observed profile (see Fig.~8). To model these absorption profiles,
we have to take into account the hyperfine structure of atoms we are
studying. In Fig. 8, we demonstrate the effects of the hyperfine
structure on the Mn~{\sc i} 5420~\textup{\AA} absorption line. In
this figure, the observed spectrum is shown as the solid black line,
the synthetic spectrum without hyperfine structure taken into
account is shown as the blue line, and the red squares show the
synthetic spectrum with hyperfine structure taken into account. The
synthetic profiles were constructed using SPECTRUM \footnote{
\url{http://www.appstate.edu/~grayro/spectrum/spectrum.html}} v2.67
code developed by Richard O. Gray (Gray \& Corbally, 1994). We note
that Fig.~8 is the same as Figure~3.1 of Gray (2010) but reproduced
using our own observed solar spectrum.

\acknowledgments We would like to thank the anonymous referee for
his/her careful reviews of the paper that include many important
points and improved significantly the clarity of the paper. HFV
would like to thank Prof. Patrick Petitjean, Prof. Robert Kurucz,
Dr. Hossein Ebadi and Dr. Hadi Rahmani for their useful discussion.
This work has been financially supported by the Research Institute
for Astronomy and Astrophysics of Maragha (RIAAM) under research
project No. 1/312. Part of this work is made in the frame of the
Iran- France Programme Hubert Curien (PHC) "Gundishapur" of
scientific exchanges initiated by the University of Tabriz and the
Institut d'Astrophysique de Paris. NSO/Kitt Peak FTS data used here
were produced by NSF/NOAO.

\makeatletter
\let\clear@thebibliography@page=\relax
\makeatother

\begin{figure*}
\hspace{-1.3cm}
\begin{tabular}{c}
\vspace{1.2cm}
\includegraphics[bb=55 487 544 656,clip=,width=1.04\hsize]{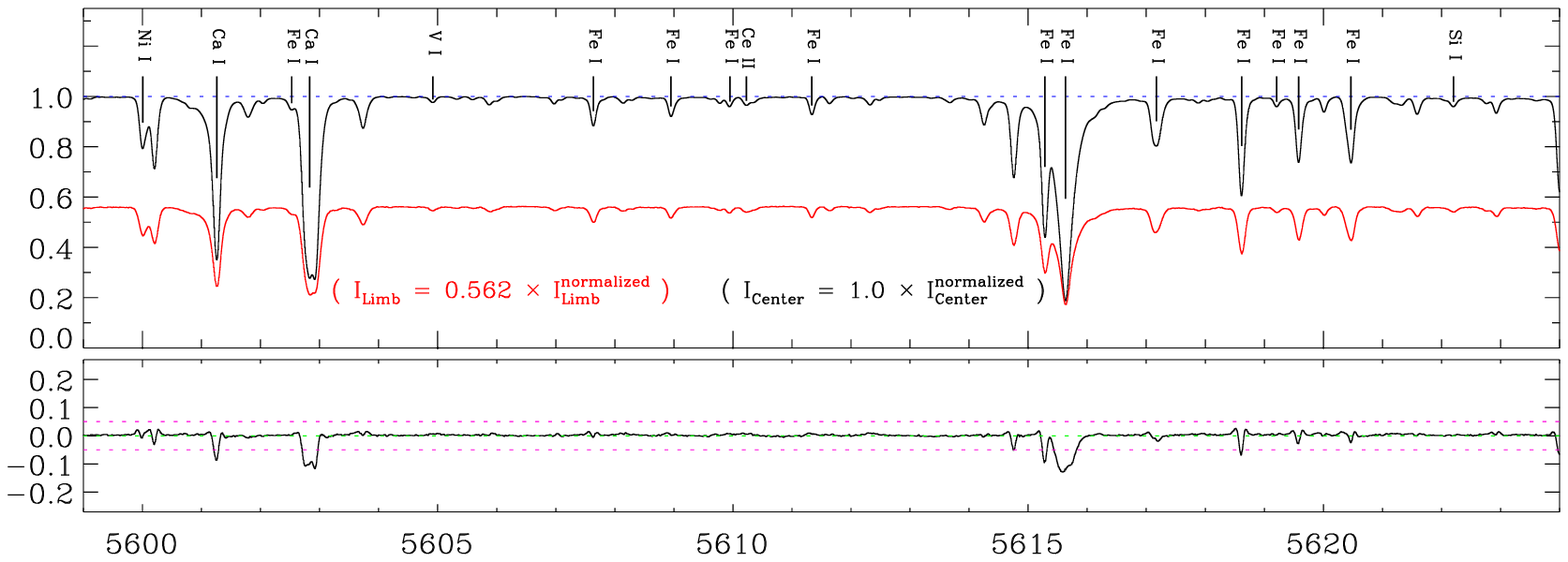}\\
\vspace{1.2cm}
\includegraphics[bb=55 487 544 656,clip=,width=1.04\hsize]{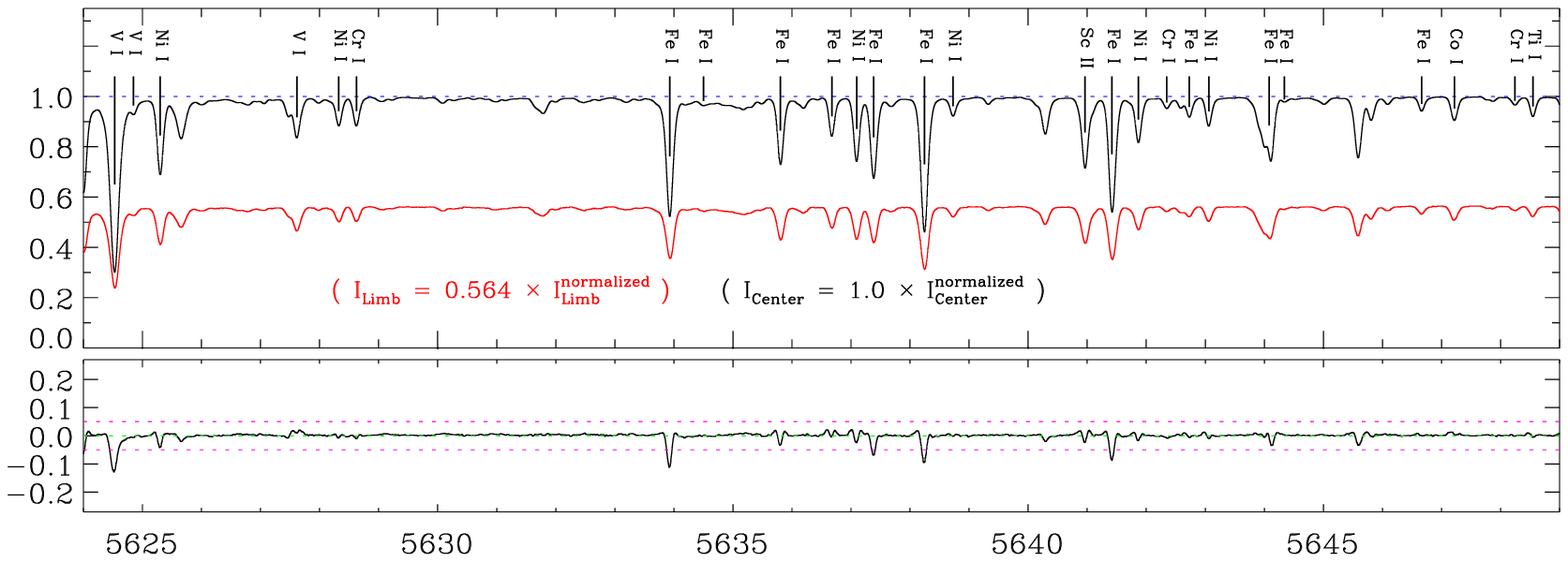}\\
\includegraphics[bb=55 487 544 656,clip=,width=1.04\hsize]{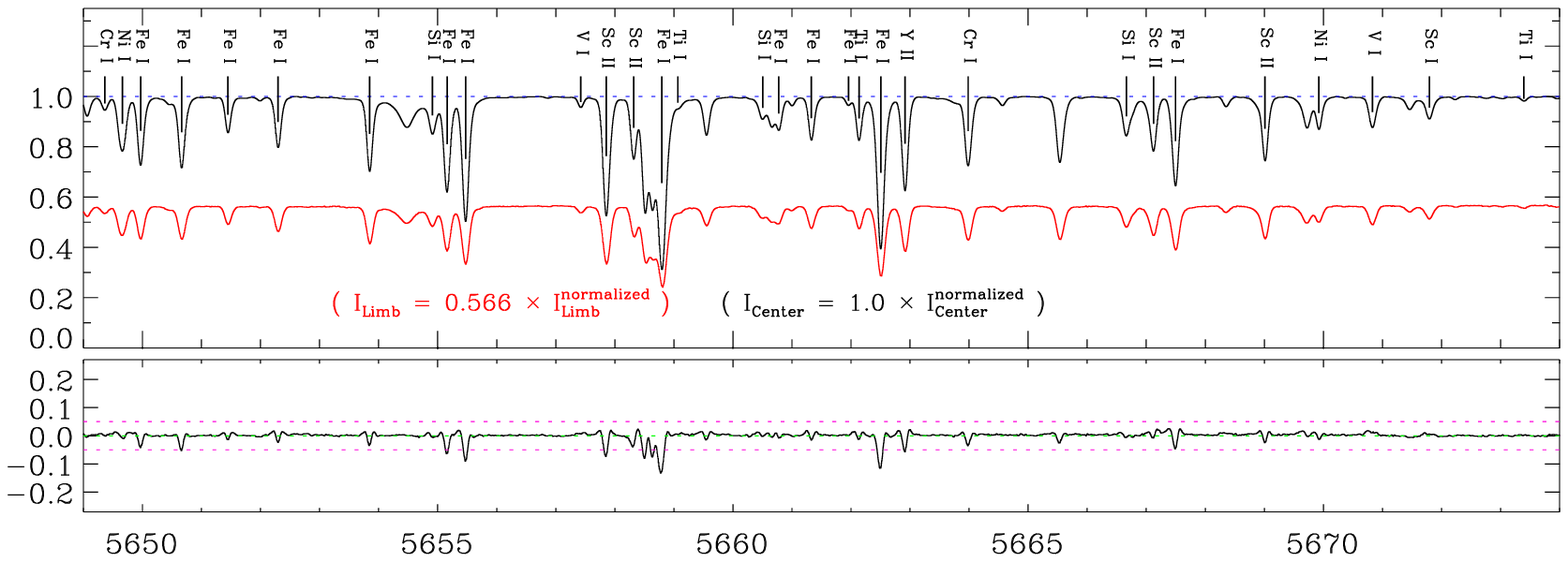}\\
\end{tabular}
 \label{landfig}
\end{figure*}

\begin{figure*}
\hspace{-1.3cm}
\begin{tabular}{c}
\vspace{1.2cm}
\includegraphics[bb=55 487 544 656,clip=,width=1.04\hsize]{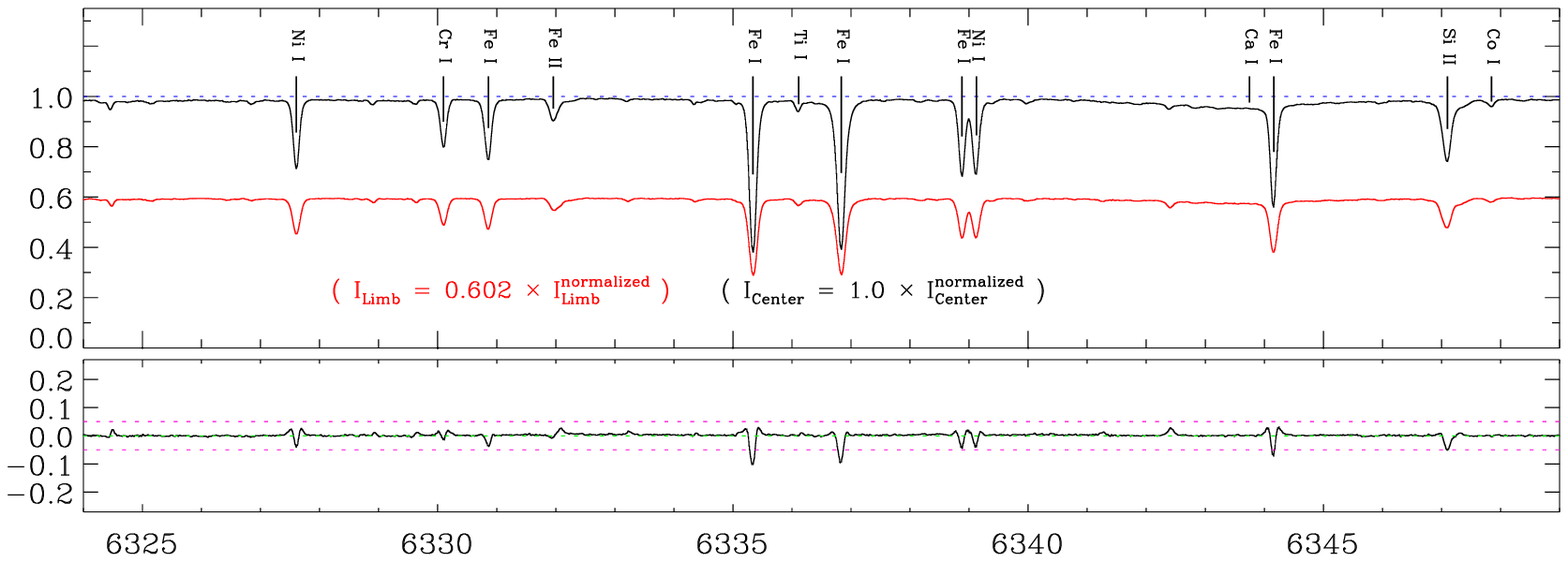}\\
\vspace{1.2cm}
\includegraphics[bb=55 487 544 656,clip=,width=1.04\hsize]{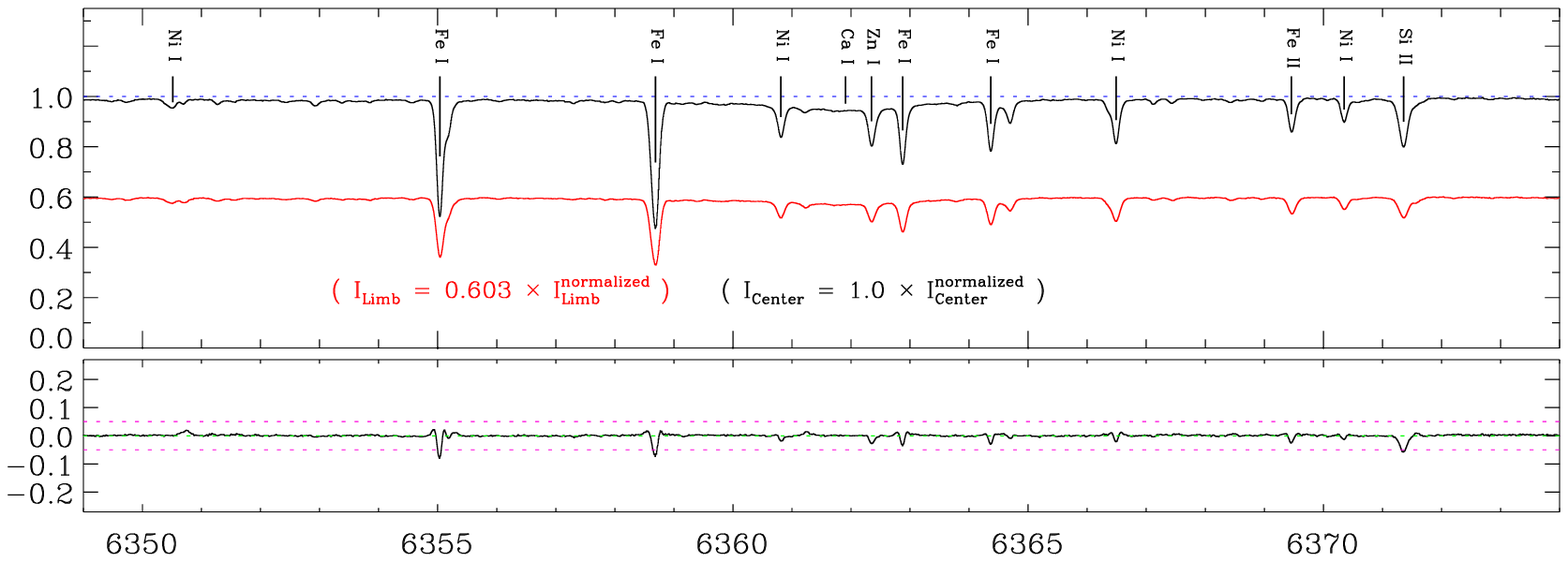}\\
\includegraphics[bb=55 487 544 656,clip=,width=1.04\hsize]{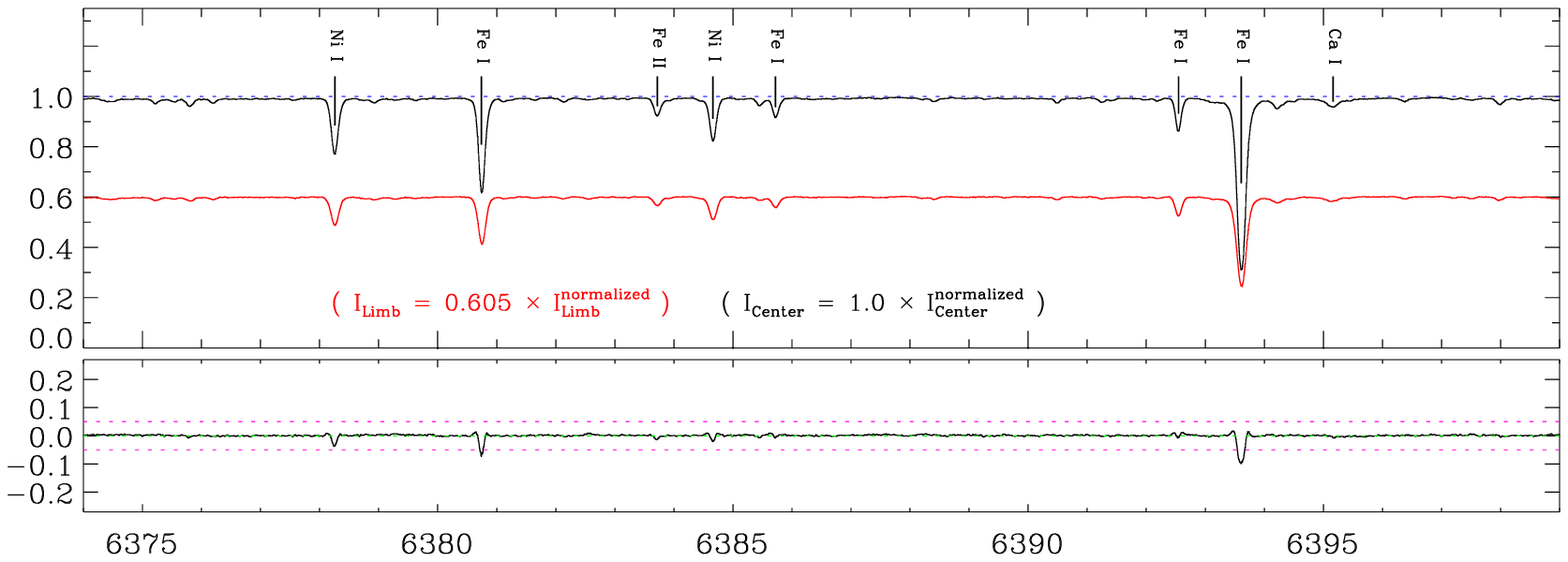}\\
\end{tabular}
 \label{landfig}
\end{figure*}

\end{document}